\documentclass[aps,prd,onecolumn,groupedaddress,showpacs,nofootinbib,amssymb]{revtex4}
\usepackage{amsmath}
\usepackage{amssymb}
\usepackage{amsfonts}
\usepackage{graphicx,bm}
\usepackage{color,amsxtra}
\usepackage{epsf}
\usepackage{enumerate}
\usepackage{hhline}
\usepackage{array}
\usepackage{tabularx}
\usepackage{colordvi}
\usepackage{color}

\newcommand{\be}{\begin{equation}}
\newcommand{\ee}{\end{equation}}
\newcommand{\bea}{\begin{eqnarray}}
\newcommand{\eea}{\end{eqnarray}}
\newcommand{\beaa}{\begin{eqnarray*}}
\newcommand{\eeaa}{\end{eqnarray*}}

\newcommand{\e}{\mathrm{e}}

\def\be{\begin{equation}}
\def\ee{\end{equation}}
\def\bea{\begin{eqnarray}}
\def\eea{\end{eqnarray}}

\def\e{\mathrm{e}}

\begin{document}

\tolerance=5000

\providecommand{\boldsymbol}[1]{\mbox{\boldmath $#1$}}

\def \lleq {\lower0.9ex\hbox{ $\buildrel < \over \sim$} ~}
\def \ggeq {\lower0.9ex\hbox{ $\buildrel > \over \sim$} ~}
\def \atridot{\stackrel{...}{a}}
\def \lam    {\Lambda}
\def\rhoc{\rho_{\rm 0c}}
\def \omx  {\Omega_X}
\def \ob  {\Omega_b}
\def \obh {\Omega_b h^2}
\def \om   {\Omega_{\rm 0m}}
\def \ol   {\Omega_{\Lambda}}
\def\oc {\Omega_{\kappa}}
\def \rb   {\bar{r}}
\def \sb   {\bar{s}}
\def \qb   {\bar{q}}
\def \beq  {\begin{equation}}
\def \eeq  {\end{equation}}
\def \ber  {\begin{eqnarray}}
\def \eer  {\end{eqnarray}}
\def \lcdm  {$\Lambda$CDM }

\title{The  evolution of gravitons  in accelerating cosmologies: the case of extended gravity}

\author{S. Capozziello$^{1,2,3}$, M. De Laurentis$^{4,5,6,2}$, S. Nojiri$^{7,8}$, S. D. Odintsov$^{9,10}$}

\affiliation{\it $^1$ Dipartimento
di Fisica``E. Pancini", Universit\`{a} di Napoli {}``Federico II''\\
$^2$INFN Sez. di Napoli, Compl. Univ. di Monte S. Angelo, Edificio G, Via Cinthia, I-80126, Napoli, Italy,\\
$^3$Gran Sasso Science Institute, Via F. Crispi 7, I-67100, L' Aquila,
	Italy,\\
$^4$Institute for Theoretical Physics, Goethe University, Max-von-Laue-Str. 1, 60438 Frankfurt, Germany,	\\
$^5$Tomsk State Pedagogical University,
634061 Tomsk, Russia\\
$^6$ Laboratory for Theoretical Cosmology,
Tomsk State University of Control Systems and Radioelectronics (TUSUR), 634050 Tomsk, Russia\\	
$^7$ Department of Physics,
Nagoya University, Nagoya 464-8602, Japan\\
$^8$ Kobayashi-Maskawa Institute for the Origin of Particles and the Universe, Nagoya University, Nagoya 464-8602, Japan\\	
$^9$ICREA, Passeig Luis Companys,
23, 08010 Barcelona, Spain, \\
$^{10}$ Institute of Space Sciences (IEEC-CSIC)
C. Can Magrans s/n, 08193 Barcelona, Spain}

\date{\today}
\begin{abstract}
We discuss the  production and evolution of cosmological gravitons  showing how the cosmological background affects their dynamics. Besides, the detection of cosmological gravitons could constitute an extremely important signature to discriminate among different cosmological models. Here we consider the cases of scalar-tensor gravity and $f(R)$ gravity where it is 
demonstrated the amplification of graviton amplitude changes  if compared with 
General Relativity. Possible observational constraints are discussed.
\end{abstract}


\maketitle

\section{Introduction}
\label{uno}
The recent discovery of gravitational waves 
\cite{Abbott:2016blz,Abbott:2016nmj} opened the doors to the so called {\it gravitational astronomy} by which a new fundamental tool to explore the universe  is at hand. Despite this amazing result,  several open issues remain to be addressed in cosmology and astrophysics, first of all the puzzle of dark side.  This issue could have a ``material" explanation by finding out new particles beyond the Standard Model or be addressed by extending the General Relativity (GR) including further degrees of freedom like scalar fields or further geometric invariants related to  curvature or torsion 
\cite{Capozziello:2011et,Nojiri:2010wj,
Peebles:2002gy,Sahni:2004ai,Bamba:2012cp,Koyama:2015vza,
Padmanabhan:2002ji,
Nojiri:2006ri,Capozziello:2007ec, manos}. 

Gravitational radiation could have a major role in this perspective because both the production and the evolution of gravitons could probe the universe at various scales \cite{Lin:2016gve}. Furthermore, the features of gravitational radiation could be an intrinsic way to test theories \cite{felix}. In fact, further gravitational modes related to the extensions of GR could be revealed both at astrophysical and cosmological scales \cite{Capozziello:2007vd,Bellucci:2008jt, Bogdanos:2009tn, novikov1,novikov2},  once suitable interferometric experiments will be available. To this end, in order to find out further polarizations of gravitational waves, more than two interferometric antennas are needed  to disentangle other modes with respect to the standard $\times$ and $+$ modes of GR \cite{Pitkin:2011yk}.

The aim of this paper is to show that the production and the evolution of gravitons can trace the evolution of cosmological models and then, in principle, distinguish among competing gravitational theories. 

Gravitons are an ideal tool because they do not interact with any form of matter and can be originated at the origin of the universe as vacuum fluctuations. Due to these characteristics, gravitons could be related to the whole cosmic history. In particular,  the behavior of cosmic accelerating/decelerating  background  is intrinsically related to the shape  of graviton amplitudes evolving with time. 

The paper is structured as follows. Sec.~\ref{conf} deals with the general problem of gravitons evolving in an expanding universe. Their conformal properties are discussed. Sec.~\ref{production} is devoted to the production of gravitons by inflationary mechanism. Such gravitons could constitute the main ingredient of the gravitational stochastic background. 
In Secs.~\ref{scalar} and \ref{tre},  scalar tensor and $f(R)$ gravity are considered respectively. The goal is to demonstrate that the behavior of gravitons can distinguish cosmological features (e.g. phantom phases 
\cite{Caldwell:1999ew} and Big Rip singularities \cite{Caldwell:2003vq}) and then discriminate among models. Discussion and conclusions are drawn in Sec.~\ref{cinque}. Here we discuss also possible experimental constraints that could be considered to implement the present approach.

\section{Gravitons in expanding universe and their conformal properties}
\label{conf}
Let us  consider the equation of motion for  gravitons in the expanding universe. 
The expansion of the universe is generated by the energy momentum tensor 
of matter acting as a source. However, the concept of source is very general and any  (material and geometric) contribution to the right-hand side of the Einstein field equations can contribute. 
The metric dependence of the energy momentum tensor is not always straightforward
and  modifications can be induced by  scalar fields, geometric invariants and any source contributing to
 the expansion of the universe (see for example
Ref.~\cite{Higuchi:2014bya}). 

Before  considering the cosmological applications, it is worth discussing conformal properties of graviton dynamics in a given expanding background. These considerations are worth in view of understanding how the evolution of  gravitons can be used to probe  different cosmological backgrounds. As we will show, the approach works for models where it is possible  to disentangle the standard fluid matter  with respect to the geometry into the Einstein field equations.  In general,  perturbing the metric means
\be
\label{H1}
g_{\mu\nu} \to g_{\mu\nu} + h_{\mu\nu}\, ,
\ee
where $|h_{\mu\nu}|\ll 1$ is the perturbation with respect to a given background $g_{\mu\nu}$.
It is straightforward to obtain  the perturbed Ricci tensor and scalar,
\begin{align}
\label{H2}
\delta R_{\mu\nu} = & \frac{1}{2} \left[ \nabla_\mu \nabla^\rho h_{\nu\rho} 
+ \nabla_\nu \nabla^\rho h_{\mu\rho} - \nabla^2 h_{\mu\nu} 
 - \nabla_\mu \nabla_\nu \left( g^{\rho\lambda} h_{\rho\lambda} \right) 
 - 2 R^{\lambda\ \rho}_{\ \nu\ \mu} h_{\lambda\rho} 
+ R^\rho_{\ \mu} h_{\mu\nu} + R^\rho_{\ \nu} h_{\rho\mu} \right] \, ,\\
\label{H3}
\delta R = & - h_{\mu\nu} R^{\mu\nu} + \nabla^\mu \nabla^\nu h_{\mu\nu} 
 - \nabla^2 \left( g^{\mu\nu} h_{\mu\nu}\right)\, .
\end{align}
We are not considering the perturbation   of  scalar fields  because the spin 
two field, i.e. the graviton, does not mix with the scalar field (spin zero) field. In this sense, the ``genuine" graviton is a tensor field propagating into a given background. However, it is worth noticing that extending GR means to take into account further degrees of freedom that can be figured out as effective scalar fields \cite{Capozziello:2007ec}.

By imposing the gauge condition
\be
\label{H4}
\nabla^\mu h_{\mu\nu} = g^{\mu\nu} h_{\mu\nu} = 0\, ,
\ee
the Einstein field equations
 \be
 \label{einstein}
 R_{\mu\nu} - \frac{1}{2} g_{\mu\nu} R = \kappa^2 T_{\mu\nu}\,,
 \ee 
assume the following perturbed form:
\be
\label{H6}
\frac{1}{2} \left[ - \nabla^2 h_{\mu\nu} 
 - 2 R^{\lambda\ \rho}_{\ \nu\ \mu} h_{\lambda\rho} 
+ R^\rho_{\ \mu} h_{\rho\nu} + R^\rho_{\ \nu} h_{\rho\mu}
 - h_{\mu\nu} R + g_{\mu\nu} R^{\rho\lambda} h_{\rho\lambda} \right] 
= \kappa^2 \delta T_{\mu\nu}\, .
\ee
Here we are assuming that  the  energy-momentum tensor perturbations, acting as a source,  lead the evolution of the gravitons.
This means that any source contributing to the r.h.s. of Eq.~\eqref{H6} 
affects the propagation of gravitons.
However, 
before considering the form of $\delta T_{\mu\nu}$, we have to  discuss  conformal properties of  gravitons showing how further degrees of freedom of gravitational field can be figured out as  further scalar fields acting as sources.

In general, gravitational waves  derive from  perturbations $h_{\mu\nu}$ of the metric $g_{\mu\nu}$
and transform as  $3$-tensors. The 
gravitational wave equation in vacuum and in the transverse-traceless gauge is
\begin{equation}
\square h_{i}^{\ j}=0\label{eq: 1}\,,
\end{equation}
where $\Box$
is the  d'Alembert operator defined as $\square\equiv(-g)^{-1/2}\partial_{\mu}(-g)^{1/2}g^{\mu\nu}\partial_{\nu}$. This equation comes directly 
from the Einstein field Eqs.~(\ref{einstein}) in the Minkowskian limit. The Latin indexes are for spatial coordinates and  the Greek
ones for spacetime coordinates. Our task is now to derive the analog of 
Eqs.~(\ref{eq: 1}) assuming a generic theory of gravity conformally related to GR.  

Assuming the conformal transformation
\begin{equation}
\label{conformal}
\widetilde{g}_{\mu\nu}=\e^{2\Phi}g_{\mu\nu} \, ,
\end{equation} 
where  $\Phi$ is the conformal scalar field,
the Ricci tensor and scalar are respectively
\begin{equation}
\widetilde{R}_{\mu\nu}=R_{\mu\nu}+2\left(\Phi_{;\mu}\Phi_{;\nu}-
g_{\mu\nu}\Phi_{;\delta}\Phi^{;\delta}-\Phi_{;\mu\nu}-\frac{1}{2}g_{\mu\nu}\Phi^{;\delta}\,_{;\delta}\right)\,,\quad
\widetilde{R}=\e^{-2\Phi}\left(R-6\square\Phi-6\Phi_{;\delta}\Phi^{;\delta}\right) \, ,
\label{eq:6}
\end{equation}
where the scalar part can be easily disentangled with respect to the tensor part.

The box 
operator transforms in the conformal metric as 
\begin{equation}
\widetilde{\square}\equiv(-\widetilde{g})^{-1/2}\partial_{\mu}(-\widetilde{g})^{1/2}\widetilde{g}^{\mu\nu}\partial_{\nu}\,,
\end{equation}
and then, applied to Eq.(\ref{conformal}), gives
\begin{equation}
\widetilde{\square}=\e^{-2\Phi}\left(\square
+2\Phi^{;\lambda}\partial_{;\lambda}\right) \, .
\label{eq:9}
\end{equation}
This means that  the gravitational wave equation in the Jordan frame  becomes
\begin{equation}
\widetilde{\square}\widetilde{h}_{i}^{\ j}=0\label{eq:7}\,,
\end{equation}
expressed in the conformal metric $\widetilde{g}_{\mu\nu}$.

Since
no scalar perturbation couples to the tensor part, we can discard the  $\delta\Phi$ contribution and,  being
$\displaystyle{\widetilde{h}_{i}^{\ j}=\widetilde{g}^{lj}\delta\widetilde{g}_{il}=\e^{-2\Phi}g^{lj} \e^{2\Phi}\delta
g_{il}=h_{i}^{\ j}}.$ In this sense, 
$h_{i}^{\ j}$ results conformally invariant.
The wave amplitude is then
\begin{equation}
\label{onda}
h(t)_{i}^{\ j}=h(t)e_{i}^{\ j}\exp(ik_{l}x^{l})\, ,
\end{equation}
where $e_{i}^{\ j}$ is the
polarization tensor for both metrics.  Time evolution is given by the scalar amplitude function $h(t)$, where we are assuming that time and space coordinates are disentangled in the evolution of gravitons. The meaning of the formula (\ref{eq:9}) is  that the background conformally  changes.
In other words, the gravitational waves   can be used to test different cosmological backgrounds.
Below, we will specify the conformal transformations for  scalar tensor and $f(R)$ gravity for Friedmann-Robertson-Walker (FRW) universes. In other words, we will  consider FRW universe in 
physical time and  in conformal time description taking into account the evolution of gravitons.
Finally, an important point has to be stressed. The
characteristic
properties of the gravitational waves are preserved
under  general
conformal transformations where $\Phi$ is a general spacetime function.
 In particular,   the transverse
property is always 
preserved by a conformal transformation if
$\widetilde
\nabla^\mu \widetilde h_{\mu\nu}=0$.

Although $\widetilde h_{\mu}^{\ \nu}=h_{\mu}^{\ \nu}$,  
even if $\nabla^\mu h_{\mu}^{\ \nu}=0$, in general  on finds 
$\widetilde \nabla^\mu \widetilde h_{\mu}^{\ \nu}\neq 0$. 
In fact we have, 
\be
\label{Cg2}
\widetilde{\nabla}^\mu \widetilde{h}_{\mu}^{\ \nu} 
= \e^{-\Phi} \nabla^\mu h_{\mu\nu} 
+ D \e^{-\Phi} g^{\mu\sigma} g^{\nu\rho} \Phi_{,\sigma} h_{\mu\rho} 
 - \e^{-\Phi} g^{\nu\rho}\Phi_{,\rho} g^{\mu\sigma} h_{\mu\sigma} \, ,
\ee
where $D$ is the dimensions of space-time. 
Then even if we impose the gauge condition (\ref{H4}), 
we obtain 
\be
\label{Cg3}
\widetilde{\nabla}^\mu \widetilde{h}_{\mu}^{\ \nu} 
= D \e^{-\Phi} g^{\mu\sigma} g^{\nu\rho} \Phi_{,\sigma} h_{\mu\rho} \, .
\ee
If we assume that the background metric and therefore $\Phi$ only 
depend on the cosmological time $t$ and we consider the perturbation 
with $h_{t\mu}=0$, the r.h.s. in (\ref{Cg3}) vanishes,  and then
\be
\label{Cg4}
\widetilde{\nabla}^\mu \widetilde{h}_{\mu}^{\ \nu} = 0\, .
\ee
Therefore the equation for the graviton is not changed by the conformal 
transformation (\ref{conformal}).  In this sense,  as we will discuss below, gravitational waves, i.e. gravitons, can be an efficient tool to probe the cosmological background.

A first systematic study in this sense is reported in 
\cite{fabris}. These authors studied  the behavior of cosmological gravitational 
waves under
conformal transformations pointing out that information carried in  Einstein's 
and  in  Jordan's  frame are different. This fact is extremely important in order 
to discriminate the physical frame.

\section{The production of cosmological gravitons}
\label{production}
Before considering specific backgrounds, it is worth discussing the mechanism of production of cosmological gravitons.
In general, the cosmic evolution of gravitons is strictly related to the  problem of their production. It is worth stressing  that  cosmological gravitons contribute to the stochastic background and can be  generated  by  several  mechanisms of cosmological and astrophysical origin  \cite{Chiba:2006jp, Maggiore:1999vm}. As a general remark, the gravitational stochastic background is essentially  due to
very energetic phenomena   in  early universe and it  is strictly related to the cosmological model.  It is possible to 
 show that   the graviton evolution    can be connected  to   given cosmological  models  assuming  that  contributions to the stochastic background come from the 
vacuum fluctuations originated at   primordial   inflation eras.  The  paradigm is that    a transition occurs between a superluminal (e.g. exponential or power law phase) and a Friedmann power-law phase. 
  Gravitons  adiabatically evolve in relation to 
damped oscillations $(\sim 1/a)$.  Here $a$ is the FRW  scale factor. The process  stops when  gravitons reach the Hubble
radius $H^{-1}$.  Here the Hubble size is the  perturbation particle horizon.   Further   fluctuations are
 negligible thanks to  the inflation. Gravitons freeze out at
 $a/k\gg H^{-1}$. As soon as  the reheating starts, gravitons  reenter the Hubble radius. Depending on the  graviton scale perturbations, such a reenter can happen  at  radiation era or at  dust era.  The Sachs-Wolfe effect on the temperature anisotropy $\bigtriangleup T/T$  is the way to detect the phenomenon \cite{Sachs:1967er}.  

The mechanism of graviton production can be outlined as follows.  Let us consider a scalar field $\Phi$
acting as  the inflaton. In order to have inflation,  it has to be  $\dot{\Phi}\ll H$, where the dot represent the derivative with respect to cosmic time.  It is worth defining  a  conformal time $d\sigma=dt/a$.  Then the conformal gravitational wave equation (\ref{eq:7}) in FRW metric, adopting the definition (\ref{eq:9}), becomes
\begin{equation}
h''+2\left(\frac{\chi'}{\chi}\right)h'+k^{2}h=0 \, ,
\label{eq:16}
\end{equation}
where $\chi=a \e^{\Phi}$ and derivation is with respect to $\sigma$. Clearly, the given cosmological model is assigned by $\Phi$.

The mechanism of production can be realized as follows.
The inflationary background is given, in general, by  $a(t)=a_{0}\exp(Ht)$ and then $\sigma=\int dt/
a=(aH)^{-1}$. This means that, inside (\ref{eq:16}), it is  $\chi'/\chi=-\sigma^{-1}$. The solution of
(\ref{eq:16}) is
\begin{equation}
h(\sigma)=k^{-3/2}\sqrt{2/k}\left[C_{1}\left(\sin k\sigma-\cos k\sigma\right)+C_{2}\left(\sin k\sigma+\cos k\sigma\right)\right]\,,
\label{eq:17}
\end{equation}
where $C_{1,2}$ are integration constants. We can distinguish between two regimes, inside and outside the Hubble radius. Inside the Hubble radius $H^{-1}$,  it is  $k\sigma\gg1.$ Supposing  that the  initial vacuum state is symmetric, this means that we have no initial graviton and then 
 only negative-frequency modes are generated. The adiabatic
behavior is then 
\begin{equation}
h=k^{1/2}\sqrt{2/\pi}\left(\frac{1}{aH}\right)C\exp(-ik\sigma)\,,
\label{eq:18}
\end{equation}
with $C$ again a constant. The change of regime is realized at the first horizon crossing $(aH=k)$. Here, the averaged amplitude $A_{h}=(k/2\pi)^{3/2}\left|h\right|$
of the perturbation can be assumed as
\be
A_{h}=\frac{1}{2\pi^{2}}C\label{eq:19}\,.
\ee
As soon as  the scale $a/k$ becomes  larger than  $H^{-1}$,
the growing  modes are  constant and result  frozen.
This situation is realized for   $-k\sigma\ll 1$ in 
Eq.~(\ref{eq:17}). The inflaton field is
$\Phi\sim 0$ at reenter. As a consequence the
amplitude $A_{h}$ of the graviton  remain the same up  to  the second
horizon crossing.  It can be observed as
  anisotropy perturbation on the  microwave background. 
In particular  $A_{h}$ is the upper limit on the cosmological  temperature perturbation which is
\be
\bigtriangleup T/T\lesssim A_{h}\,,
\label{upper}
\ee
and this means that  other effects can bring contributions  to the background anisotropy
\cite{key-19}. 
In  Eq.~(\ref{upper}),  the 
important  quantity is  $C$, the amplitude in Eq.~(\ref{eq:18}). It is  conserved up to the reenter. 
It strictly  depends on  inflation that
produce perturbations by means of zero-point energy fluctuations. 

Specifically the production 
mechanism depends on the  specific theory of gravity that gives the background and then the inflationary behavior. From an observational point of view, 
$(\bigtriangleup T/T)$, through $A_h$,   is  a further constraint to
select cosmological models and then to select possible modified gravity theories \cite{Ade:2015rim}.

Coming into details, one can explicitly show how the graviton amplitude is related to the background and the field sourcing the inflation. Let us take into account a single graviton described as 
 a  monochromatic wave.  Its zero-point amplitude is
derived from the commutation relation:
\begin{equation}
\left[h(t,x),\,\pi_{h}(t,y)\right]=i\delta^{3}(x-y)\,,
\label{eq:20}
\end{equation}
calculated at  time $t$.  Here the amplitude $h$ has the role of the
field and $\pi_{h}$ is its conjugate momentum. One can write an effective interaction
 Lagrangian for $h$ as
\begin{equation}
\widetilde{\mathcal{L}}
=\frac{1}{2}\sqrt{-\widetilde{g}}\widetilde{g}^{\mu\nu}h_{;\mu}h{}_{;\nu} \, ,
\label{eq:21}
\end{equation}
in the conformal  metric $\widetilde{g}_{\mu\nu}$. 
From this Lagrangian, the conjugate momentum is 
\begin{equation}
\pi_{h}=\frac{\partial\widetilde{\mathcal{L}}}{\partial\dot{h}}
= \e^{2\Phi}a^{3}\dot{h}\label{eq:22}\,.
\end{equation}
Immediately, the formal Eq.~(\ref{eq:20}) becomes
\begin{equation}
\left[h(t,x),\,\dot{h}(t,y)\right]=i\frac{\delta^{3}(x-y)}{a^{3}\e^{2\Phi}} \, .
\label{eq:23}
\end{equation}
Functions $h$ and $\dot{h}$ are  expandable  through
creation and annihilation operators, i.e.  
\begin{equation}
h(t,x)=\frac{1}{(2\pi)^{3/2}}\int d^{3}k\left[h(t) \e^{-ikx}+h^{*}(t)e^{+ikx}\right]
\, ,\quad
\dot{h}(t,x)=\frac{1}{(2\pi)^{3/2}}\int d^{3}k\left[\dot{h}(t) \e^{-ikx}+\dot{h}^{*}(t) \e^{+ikx}\right]\,.
\label{eq:25}
\end{equation}
Taking into account  the conformal time,  the commutation relations are 
\begin{equation}
\left[hh'^{*}-h^{*}h'\right]=\frac{i(2\pi)^{3}}{a^{3} \e^{2\Phi}} \, .
\label{eq:26}
\end{equation}

Considering the solution (\ref{eq:18}) and (\ref{eq:19}) of cosmological equation (\ref{eq:16}), we obtain
$C=\sqrt{2}\pi^{2}H \e^{-\Phi}$ where $H$ and $\Phi$ are calculated
at the first horizon-crossing. Explicitly we obtain
\begin{equation}
A_{h}=\frac{\sqrt{2}}{2}H \e^{-\Phi} \, ,
\label{eq:27}
\end{equation}
which means that the amplitude of gravitons produced during inflation
directly depends on the cosmological model (see also \cite{cordac,mpla} for a detailed discussion on this point).  Clearly, $\e^{\Phi}=1$ for GR.
Below, we shall consider two relevant cases, scalar-tensor and $f(R)$ gravity, showing that graviton  evolution can be a relevant tool to discriminate  features of  cosmological models.

\section{The scalar tensor case}
\label{scalar}
Let us develop the  above considerations  for some specific  theories of gravity. First of all, we have to specify  
 the explicit form of $\delta T_{\mu\nu}$.  Assuming the scalar field model  in \cite{Nojiri:2005pu} whose action is given by
\be
\label{H7}
S_\phi = \int d^4 x \sqrt{-g} \mathcal{L}_\phi\, , \quad 
\mathcal{L}_\phi =  - \frac{1}{2} \omega(\phi) \partial_\mu \phi 
\partial^\mu \phi - V(\phi) \, ,
\ee
we find 
\be
\label{H8}
T_{\mu\nu} = - \omega(\phi) \partial_\mu \phi \partial_\nu \phi 
+ g_{\mu\nu} \mathcal{L}_\phi\, ,
\ee
and therefore 
\be
\label{H9}
\delta T_{\mu\nu} = h_{\mu\nu} \mathcal{L}_\phi 
+ \frac{1}{2} g_{\mu\nu} \omega(\phi) \partial^\rho \phi 
\partial^\lambda \phi h_{\rho\lambda}\, ,
\ee 
up to first order in perturbations.
We are interested  in the evolution of tensor gravitons so we can consider only the spatial component of 
 $h_{\mu\nu}$, that is $h_{ij}$.
 
Assuming a FRW spatially flat metric
\be
\label{FRWmetric}
ds^2 = - dt^2 + a(t)^2 \sum_{i=1,2,3} \left( dx^i \right)^2 \, ,
\ee
and  $\phi=t$ in  (\ref{H9}),
the FRW equations are 
\be
\label{LB1}
\frac{3}{\kappa^2} \frac{{\dot a}^2}{a^2} = \frac{\omega}{2} + V\, ,\quad 
 - \frac{1}{\kappa^2} \left( 2\frac{\ddot{a}}{a} + \frac{\dot{a}^2}{a^2}
\right)
= \frac{\omega}{2} - V\, .
\ee
We find 
\be
\label{LB2}
\omega = - \frac{2}{\kappa^2} \left( \frac{\ddot{a}}{a} - \frac{\dot{a}^2}{a^2}
\right) \, , \quad 
V = \frac{1}{\kappa^2} \left( \frac{\ddot{a}}{a} 
+ 2 \frac{\dot{a}^2}{a^2} \right)
\, ,
\ee
for the kinetic term and the scalar field potential expressed as functions of the scale factor $a(t)$ and its derivatives.
By using (\ref{H6}), (\ref{H9}), (\ref{LB2}), and $\phi=t$, 
we find the evolution equation of graviton:
\be
\label{LB3}
0=\left(2\frac{\ddot{a}}{a} + 4 \frac{{\dot a}^2}{a^2}
+\frac{\dot{a}}{a}\partial_t-\partial_t^2 
+\frac{\bigtriangleup}{a^2}\right)h_{ij}\, ,
\ee
which clearly depends on the cosmological background.
We have to  compare Eq.~(\ref{LB3}) with the expression in case of 
$\delta T_{\mu\nu} =0$. We obtain
\be
\label{LB3B}
0=\left(6\frac{\ddot{a}}{a} + 6 \frac{{\dot a}^2}{a^2}
+\frac{\dot{a}}{a}\partial_t-\partial_t^2 
+\frac{\bigtriangleup}{a^2}\right)h_{ij}\, ,
\ee
where the different coefficients specify the role of scalar field $\phi$,  and then the matter source, in the evolution. Clearly, Eqs.~(\ref{LB3}) and (\ref{LB3B}) are conformally related according to the discussion in 
Sec.~\ref{conf}.

Let us  now investigate the solution of Eq.~(\ref{LB3}). 
By rewriting Eq.~(\ref{LB3}) as follows, 
\be
\label{grv1}
0 = - a^3 \partial_t \left( a^3 \partial_t \left( a^{-2} h_{ij} \right) \right) 
+ 4 a^4 {\dot a}^2 \left( a^{-2} h_{ij} \right)
+ a^4 \bigtriangleup \left( a^{-2} h_{ij}\right) \, ,
\ee
and assuming $h_{ij}$  given by the above spatially plane wave (\ref{onda}), one has 
\be
\label{grv2}
h_{ij} \left(\bm{x}, t \right) = \e^{i\bm{k}\cdot \bm{x}} a(t)^2 \hat h_{ij}(t)\, . 
\ee
We can define a new conformal time coordinate $\tau$ as\footnote{It is worth noticing that the conformal time $\sigma$ adopted above is useful for the scaling properties of the graviton while $\tau$ is considering the volume invariance related to $a^3$. The time $\tau$ can be always related to $\sigma$.} 
\be
\label{grv3}
d\tau = a(t)^{-3}dt \, ,
\ee
and then we obtain 
\be
\label{grv4}
0 = \frac{d^2 \hat h_{ij}}{d\tau^2} 
+ 4 a^4 {\dot a}^2 \hat h_{ij}
+ k^2 a^4 \hat h_{ij}\, .
\ee
Here $k^2 \equiv \bm{k}\cdot \bm{k}$ is the wave number of Eq.~(\ref{eq:16}). 

To study the evolution of gravitons after the   inflationary production,  let us consider a power-law behavior for the scale factor of the universe, that is 
\be
\label{grv5}
a(t) = \left( \frac{t}{t_0} \right)^\alpha \, ,
\ee
with  $t_0$ and $\alpha$ real constants. Depending on the value of $\alpha$, Eq.~(\ref{grv5}) involves a power law (superluminal) inflation  $(\alpha \geq 1)$, a Friedmannian (subluminal) evolution $(0<\alpha < 1)$, and a pole-like (phantom) behavior $(\alpha<0)$.

The universe given by (\ref{grv5}) can be realized by the scalar-tensor model in 
(\ref{H7}) by substituting (\ref{grv5}) into the expressions for  $\omega(\phi)$ and $V(\phi)$ given in (\ref{LB2}), that is
\be
\label{LB2power}
\omega (\phi)  = \frac{2\alpha }{\kappa^2 t_0^2 \phi^2} \, , \quad 
V(\phi) = \frac{3\alpha^2 - \alpha}{\kappa^2 t_0^2 \phi^2} \, .
\ee
Eq.~(\ref{grv3}) tells us that, if $\alpha\neq \frac{1}{3}$, the conformal time is
\be
\label{grv6}
\tau = \frac{t_0}{1 - 3 \alpha} \left( \frac{t}{t_0} \right)^{1 - 3 \alpha} \, .
\ee
Therefore $t\to 0$ corresponds to $\tau \to 0$ if $\alpha<\frac{1}{3}$ and $\tau \to - \infty$ 
if $\alpha>\frac{1}{3}$. 
On the other hand $t \to + \infty$ corresponds to $\tau \to +\infty$ 
if $\alpha<\frac{1}{3}$ and $\tau \to 0$ if $\alpha>\frac{1}{3}$. 
It is worth noticing that $\alpha>\frac{1}{3}$ corresponds to the accelerating 
expansion of the universe and $\alpha<\frac{1}{3}$ to the decelerating expansion. From a physical point of view,   $a(t)\sim t^{1/3}$ represents a   stiff-matter behavior \cite{Zeldovich:1972zz}.
Furthermore we have that 
\be
\label{grv7}
k^2 a^4 = k^2 \left( \frac{\tau}{\tau_0} \right)^{\frac{4\alpha}{1 - 3\alpha}}\, , \quad 
4 a^4 {\dot a}^2 = \frac{4\alpha^2}{t_0^2} \left( \frac{\tau}{\tau_0} \right)^{-2}\, .
\ee
Here
$\tau_0 \equiv \frac{t_0}{1 - 3\alpha}$. 
Because $\frac{4\alpha}{1 - 3\alpha}  - \left(-2\right) 
= \frac{2 \left( 1 - \alpha\right)}{1 - 3\alpha}$, if $\alpha<\frac{1}{3}$ or 
$\alpha>1$, the first term in (\ref{grv7}) dominates 
when $\left| \tau \right| \to \infty$ and the second term dominates when $\tau\to 0$. 
On the other hand, if $\frac{1}{3}<\alpha<1$, the second term in 
(\ref{grv7}) dominates when $\left| \tau \right| \to \infty$ and 
the first term dominates when $\tau\to 0$. In other words, the graviton behavior can discriminate among the various cosmological regimes.
Let us  now assume $\hat h_{ij}$ behaves asymptotically  as
\be
\label{grv7b}
\hat h_{ij} \propto \e^{\beta \left( \frac{\tau}{\tau_0} \right)^\gamma }\, .
\ee
Then
\be
\label{grv8}
\frac{d^2 \hat h_{ij}}{d\tau^2} = \frac{1}{\tau_0^2} \left[
\beta \gamma \left( \gamma - 1 \right) \left( \frac{\tau}{\tau_0} \right)^{\gamma - 2} 
+ \beta^2 \gamma^2 \left( \frac{\tau}{\tau_0} \right)^{ 2 \left( \gamma - 1 \right) } 
\right] 
\hat h_{ij}\, .
\ee
Because $\gamma - 2 - 2 \left( \gamma - 1 \right) = - \gamma$, 
if $\gamma>0$, the first term in the r.h.s. of (\ref{grv8}) dominates when  $\tau \to 0$ and 
the second term dominates when $\left| \tau \right| \to \infty$.
If $\gamma<0$, the second term in (\ref{grv8}) dominates when  $\tau \to 0$ and 
the first term dominates when $\left| \tau \right| \to \infty$. 

When the first term in the r.h.s. of (\ref{grv8}) dominate, 
by using (\ref{grv4}) and (\ref{grv7}), 
we find 
\be
\label{grv9}
\gamma = \frac{2 \left( 1 - \alpha\right)}{1 - 3\alpha}\, , \quad 
\beta =  - \frac{ \left( 1 - 3 \alpha \right)^2 k^2}{ 2 \left( 1 - \alpha\right) 
\left( 1 + \alpha \right)}\, .
\ee
Then combining (\ref{grv6}), (\ref{grv7b}), and (\ref{grv9}), we find 
\be
\label{grv7c}
\hat h_{ij} \propto 
\e^{- \frac{ \left( 1 - 3 \alpha \right)^2 k^2}{ 2 \left( 1 - \alpha\right) 
\left( 1 + \alpha \right)} \left( \frac{t}{t_0} \right)^{2\left( 1 - \alpha \right) }}\, ,
\ee
which is not a plane wave. 

On the other hand, when  the second term in 
(\ref{grv8}) dominates, we find 
\be
\label{grv10}
\gamma = \frac{1 - \alpha}{1 - 3\alpha} \, , \quad 
\beta = \pm i \left(\frac{1 - 3\alpha}{1 - \alpha}\right) k \, ,
\ee
which corresponding to the oscillating mode. 
By using (\ref{grv6}) and (\ref{grv7b}), and (\ref{grv10}), we obtain 
\be
\label{grv7d}
\hat h_{ij} \propto 
\e^{\pm i \left(\frac{1 - 3\alpha}{1 - \alpha}\right) k
\left( \frac{t}{t_0} \right)^{1 - \alpha }}\, .
\ee
The obtained expression is oscillating but it is not a simple plane wave except in the limit 
of $\alpha\to 0$. 
In the above two cases, 
we find that $\gamma>0$ when $0<\alpha<\frac{1}{3}$ or $\alpha>1$
and $\gamma<0$ when $\frac{1}{3}<\alpha<1$. 

Furthermore when the second term (\ref{grv7}) and the first term in (\ref{grv8}) dominate, 
if we assume (\ref{grv10c}), we find $\gamma=0$, 
that is, $\hat h_{ij}$ becomes a constant, which may tell that we need an ansatz different from 
(\ref{grv7b}) like
\be
\label{grv10c}
\hat h_{ij} \propto \left( \frac{\tau}{\tau_0} \right)^\eta \, ,
\ee
with a constant $\eta$ and we find 
\be
\label{grv10d}
0 = \frac{\eta \left( \eta - 1 \right)}{\tau_0^2} + \frac{4\alpha^2}{t_0^2} \, ,
\ee
that is 
\be
\label{grv9b}
\eta = \frac{ 1 \pm \sqrt{ 1 - 16 \alpha^2 \frac{\tau_0^2}{t_0^2}}}{2}
= \frac{ 1 \pm \sqrt{ 1 - \frac{16\alpha^2}{ \left( 1 - 3 \alpha \right)^2 }}}{2}\, .
\ee
In terms of the cosmic time $t$, Eq.~(\ref{grv10c}) can be written as
\be
\label{grv9c}
\hat h_{ij} \sim \left( \frac{t}{t_0} \right)^{\eta\left( 1 - 3 \alpha \right)}\, .
\ee
In case that the second  terms in (\ref{grv7}) and (\ref{grv8}) dominate,  if we assume the behavior (\ref{grv10c}), 
we find $\gamma=0$, again, which tells us that $\hat h_{ij}$ is given by (\ref{grv10c}) and we re-obtain 
(\ref{grv10d}) and (\ref{grv9b}). 
It is worth noticing that $\eta$ becomes imaginary if $\alpha< -1 $ or $\alpha>\frac{1}{7}$ but the real part of $\eta$ is always positive  and, therefore, we find that when $\tau \to 0$, $\hat h_{ij} \to 0$ and when $\left|\tau\right| \to \infty$, 
$\hat h_{ij}$ diverges.   

We can summarize the above considerations as follows.
\begin{enumerate}
\item When $t\to 0$,
\begin{enumerate}
\item $\alpha>1$. In this case,  $t\to 0$ corresponds to $\tau\to -\infty$ as written after 
Eq.~(\ref{grv6}). Then the first term in (\ref{grv7}) dominates. Furthermore because $\alpha>1$ tells 
$\gamma>0$, the second term in the r.h.s. of (\ref{grv8}) dominates and therefore the solution oscillates but 
the absolute value of $\hat h_{ij}$ is finite.  
\item $\frac{1}{3}<\alpha<1$. Even in this case, $t\to 0$ corresponds to $\tau\to -\infty$. 
Because the second term in (\ref{grv7}) dominates, the solution is given by (\ref{grv10c}) and diverges as in (\ref{grv9c}). 
\item $0<\alpha<\frac{1}{3}$ case. Because $t\to 0$ corresponds to $\tau\to 0$, the   
first term in (\ref{grv7}) dominates and  $\gamma>0$, which tells that the first term in the r.h.s. of  
(\ref{grv8}) dominates and by using Eq.~(\ref{grv7c}), we find $\hat h_{ij}$ goes to a finite 
value. 
\end{enumerate}
\item When $t\to +\infty$, 
\begin{enumerate}
\item $\alpha>1$. Because $t\to +\infty$ corresponds to $\tau\to 0$, the second term in (\ref{grv7}) 
dominates and the solution is given by (\ref{grv9c}), which decreases for large $t$. 
\item $\frac{1}{3}<\alpha<1$. Because $t\to +\infty$ corresponds to $\tau\to 0$, again, the   
first term in (\ref{grv7}) dominates. Then the second term in the r.h.s. of (\ref{grv8}) dominates, 
the solution oscillates as in (\ref{grv7d}) and the absolute value of $\hat h_{ij}$ is finite. 
\item $0<\alpha<\frac{1}{3}$ case. Because $\gamma>0$ and $t\to +\infty$ corresponds to 
$\tau\to +\infty$, the solution is given by (\ref{grv7d}). Therefore the solution begins to oscillate but the 
absolute value of $\hat h_{ij}$ can be finite. 
\end{enumerate}
\end{enumerate}
In summary,  we find the enhacement of the gravitational wave when $\alpha>0$ in late time.  Such a behavior could be interesting from an observational point of view. 
Because $\hat h_{ij}$ could be always finite, Eq.~(\ref{grv2}) tells that the gravitational wave $h_{ij}$ 
grows up by the factor $a(t)^2$. 

For  comparison, we can consider the model with $\delta T_{\mu\nu}=0$ in (\ref{LB3B}), that is when the scalar field does not act as a source for the gravitational equations. 
Then by using (\ref{grv2}) and (\ref{grv3}), instead of Eq.~(\ref{grv4}), we obtain 
\be
\label{grv4A}
0 = \frac{d^2 \hat h_{ij}}{d\tau^2} 
+ \left( 6 a^4 {\dot a}^2 + 4 a^6 \ddot a \right) \hat h_{ij}
+ k^4 a^4 \hat h_{ij}\, .
\ee
In the case of the power-law scale factor in (\ref{grv5}), instead of (\ref{grv7}), we find 
\be
\label{grv7R}
k^2 a^4 = k^2 \left( \frac{\tau}{\tau_0} \right)^{\frac{4\alpha}{1 - 3\alpha}}\, , \quad 
6 a^4 {\dot a}^2 + 4 a^6 \ddot a = \frac{ 10 \alpha^2 - 4 \alpha }{t_0^2} 
\left( \frac{\tau}{\tau_0} \right)^{-2}\, .
\ee
Then as in (\ref{grv7}), if $\alpha<\frac{1}{3}$ or 
$\alpha>1$, the first term in (\ref{grv7}) dominates 
when $\left| \tau \right| \to \infty$ and the second term dominates when $\tau\to 0$. 
And furthermore, if $\frac{1}{3}<\alpha<1$, the second term in 
(\ref{grv7}) dominates when $\left| \tau \right| \to \infty$ and the first term dominates when $\tau\to 0$.
The first term in (\ref{grv7R}) is not changed from the first term in (\ref{grv7}), there are some modifications 
if $\alpha<\frac{1}{3}$ or $\alpha>1$ and $\tau\to 0$, or if $\frac{1}{3}<\alpha<1$ and 
$\left| \tau \right| \to \infty$. 
In these cases, Eq.~(\ref{grv10d}) is changed as 
\be
\label{grv10dR}
0 = \frac{\eta \left( \eta - 1 \right)}{\tau_0^2} + \frac{10 \alpha^2 - 4 \alpha}{t_0^2} \, ,
\ee
and therefore we obtain 
\be
\label{grv9bR}
\eta 
= \eta_\pm \equiv \frac{ 1 \pm \sqrt{ 1 - 4 \left( 10 \alpha^2 - 4 \alpha\right) \frac{\tau_0^2}{t_0^2}}}{2}
= \frac{ 1 \pm \sqrt{ 1 - 4\frac{10 \alpha^2 - 4 \alpha }{ \left( 1 - 3 \alpha \right)^2 }}}{2}\, .
\ee
In this case, the real part of $\eta$ is not always positive but if $0<\alpha<\frac{2}{5}$, $\eta_-$ 
is negative. In conclusions, we can say that the cosmological evolution of the graviton depends on the cosmic background. Furthermore, the behavior of the graviton could be used to discriminate between the presence or not of a scalar field acting as a source in the field equations.

\section{The $f(R)$ case}
\label{tre}
Similar considerations can be developed also in the case of $f(R)$ since the further degrees of freedom coming from the extension of GR (i.e. $f(R)\neq R$) can be always figured out as a further scalar field \cite{Capozziello:2011et,Nojiri:2010wj}.
For the sake of simplicity, let us assume a 
 power-law $f(R)$ function of the form  
\be
\label{FR}
f(R) \sim R^m\, ,
\ee
and a power-law scale factor (\ref{grv5}). The exponents $\alpha$ and $m$ are related as follows
\be
\label{grv11}
\alpha = - \frac{(m-1)(2m-1)}{m-2}\, .
\ee
We have to  note that there are extrema for $\alpha=- 5 \pm 2\sqrt{6}$ at 
$m= 2 \mp \sqrt{\frac{3}{2}}$. 
In $f(R)$ gravity, the above conformal  transformation (\ref{conformal}) is 
\be
\label{grv12}
g_{\mu\nu} = \frac{1}{f'(R)} g_{\mathrm{E}\, \mu\nu}\, , 
\quad \mbox{being}\quad f'(R)=\e^{2\Phi}\,,
\ee
mapping the Jordan frame into to the Einstein frame. 
In the Einstein frame, the graviton modes and the scalar mode can be separated and we can easily 
find the equation of graviton (\ref{LB3}). In the Einstein frame, the further degrees of freedom of $f(R)$ gravity can be recast as a scalar  source in $T_{\mu\nu}$ and separated with respect to the evolution of the tensorial graviton. Below we will discuss the relative contributions of tensor and scalar modes in the stochastic background.

Considering again the power-law scale factor  (\ref{grv5}), the scalar curvature is proportional to ${\displaystyle \frac{1}{t^2}}$, 
then   $f'(R)$ can be written as  
\be
\label{grv13}
f'(R) \sim f_0 \left( \frac{t}{t_0} \right)^{-2 (m - 1)}\, .
\ee
The  metric (\ref{grv12}) in the Einstein frame is given by
\be
\label{grv14}
ds^2_E = f_0 \left( \frac{t}{t_0} \right)^{-2 (m - 1)} \left[ - dt^2 
+ \left( \frac{t}{t_0} \right)^{ - \frac{2(m-1)(2m-1)}{m-2}}
 \sum_{i=1,2,3} \left( dx^i \right)^2 
 \right]\, .
\ee
A time coordinate $t_E$ in the Einstein frame  can be defined as follows
\be
\label{grv15}
t_E = t_{E0}\left( \frac{t}{t_0} \right)^{ 2- m } \, .
\ee
Here
\be
\label{grv16}
t_{E0} \equiv \frac{ \sqrt{f_0 t_0 }}{ 2 - m} \, .
\ee
It is worth  noticing  that $t\to 0$ corresponds to $t_E\to 0$, if $m<2$ but to $t_E \to \infty$, if $m>2$ and 
$t\to \infty$ corresponds to $t_E\to \infty$, if $m<2$ but to $t_E \to 0$ if $m>2$. 
Then the metric in the Einstein frame is given by 
\be
\label{grv17}
ds^2_E = - dt_E^2 
+ \left( \frac{t_E}{t_{E0}}\right)^{6 \frac{(m-1)^2}{(m-2)^2}}
 \sum_{i=1,2,3} \left( dx^i \right)^2 \, .
\ee
Then $\tilde\alpha \equiv 3 \frac{(m-1)^2}{(m-2)^2}$ can be identified with $\alpha$ in (\ref{grv5}) 
in the Einstein frame, which is different from $\alpha$ in the Jordan frame. 

Since, in the Einstein frame,  we can write the action in the scalar-tensor form, we can use the previous 
arguments. 
We find the following correspondence: 1) $\tilde\alpha=0\, \Leftrightarrow\, \alpha=0\, , \ (m=1)$; 
2)  $\tilde\alpha=+\infty\, \Leftrightarrow\, \alpha=\pm \infty \, , \ (m=2)$; 
3) $\tilde\alpha = \frac{1}{3}\, \Leftrightarrow\, \alpha = 0\, , \ \left(m= \frac{1}{2} \right)$ 
or $\alpha = 1 , \ \left( m= \frac{5}{4} \right)$; 
4) $\tilde \alpha = 1\, \Leftrightarrow\, \alpha = 1\, , \ \left( m = \frac{1 \pm \sqrt{3}}{2} \right)$. 
It is important to stress that there is no one-to-one correspondence between $\tilde\alpha$ and $m$ or $\alpha$. 
Then by using the results about the scalar tensor theory, we find:
\begin{enumerate}
\item When $t\to 0$,
\begin{enumerate}
\item $m<\frac{1 - \sqrt{3}}{2}$ $\left( \alpha>1\, , \ 1 < \tilde \alpha <3 \right)$: 
The solution $\hat h_{ij}$ oscillates but its absolute value is finite.  
\item $\frac{1 - \sqrt{3}}{2} < m < \frac{1}{2}$ 
$\left( 0<\alpha<1\, , \ \frac{1}{3} < \tilde \alpha < 1 \right)$:
The solution is given by (\ref{grv10c}) and diverges as (\ref{grv9c}). 
\item $\frac{1}{2}< m < \frac{5}{4}$ 
$\left( - 5 + 2\sqrt{6} \leq \alpha<\frac{1}{3} \, , \  0< \tilde \alpha < \frac{1}{3} \right)$:
The solution $\hat h_{ij}$ goes to a finite value.
\item $\frac{5}{4}< m < \frac{ 1 + \sqrt{3}}{2}$ 
$\left( \frac{1}{3} <\alpha<1\, , \  \frac{1}{3} < \tilde \alpha < 1 \right)$:
The solution is given by (\ref{grv10c}) and diverges as (\ref{grv9c}). 
\item $\frac{ 1 + \sqrt{3}}{2} < m < 2$ $\left( \alpha>1\, , \  \tilde \alpha > 1 \right)$:
The solution $\hat h_{ij}$ oscillates but its absolute value is finite.  
\item $m > 2$ $\left( \alpha \leq - 5 - 2\sqrt{6}\, , \  \tilde \alpha > 3 \right)$:
We note that this case corresponds to $t_E\to \infty$. 
Then the solution decreases for small $t$.
\end{enumerate}
\item When $t\to +\infty$, 
\begin{enumerate}
\item $m<\frac{1 - \sqrt{3}}{2}$ $\left( \alpha>1\, , \ 1 < \tilde \alpha <3 \right)$: 
The solution decreases for large $t$.
\item $\frac{1 - \sqrt{3}}{2} < m < \frac{1}{2}$ 
$\left( 0<\alpha<1\, , \ \frac{1}{3} < \tilde \alpha < 1 \right)$:
The solution oscillates and the absolute value of $\hat h_{ij}$ is finite.
\item $\frac{1}{2}< m < \frac{5}{4}$ 
$\left( - 5 + 2\sqrt{6} \leq \alpha<\frac{1}{3} \, , \  0< \tilde \alpha < \frac{1}{3} \right)$:
The solution oscillates and the absolute value of $\hat h_{ij}$ is finite.
\item $\frac{5}{4}< m < \frac{ 1 + \sqrt{3}}{2}$ 
$\left( \frac{1}{3} <\alpha<1\, , \  \frac{1}{3} < \tilde \alpha < 1 \right)$:
The solution oscillates and the absolute value of $\hat h_{ij}$ is finite.
\item $\frac{ 1 + \sqrt{3}}{2} < m < 2$ $\left( \alpha>1\, , \  \tilde \alpha > 1 \right)$:
The solution decreases for large $t$.
\item $m > 2$ $\left( \alpha \leq - 5 - 2\sqrt{6}\, , \  \tilde \alpha > 3 \right)$:
This case corresponds to $t_E\to 0$ case.
Therefore the solution oscillates but the absolute value of $\hat h_{ij}$ is finite.  
\end{enumerate}
\end{enumerate}
Then we find that when  $m<\frac{1 - \sqrt{3}}{2}$ or $2> m > \frac{ 1 + \sqrt{3}}{2}$, 
the amplitude of the solution is suppressed in the late time.  
For $m>2$ or $\frac{1}{2}<m<1$,  $\alpha$ 
assumes negative values and therefore the universe undergoes a phantom evolution. 
It is worth noticing that the phantom universe does not appear in the case of  scalar tensor 
theory.  
For phantom universe, $t\to \infty$ corresponds to the infinite past
by changing the direction of time 
and $t\to 0$ corresponds to the Big Rip singularity. 
Near the Big Rip singularity, in case $m>2$,  
$\hat h_{ij}$ decreases; for  $\frac{1}{2}<m<1$, $\hat h_{ij}$ is finite and not oscillating 
near the Big Rip singularity for $t\to 0$.  

In the case $\alpha>0$,  corresponding to the quintessence universe, $\hat h_{ij}$ decreases
for both  scalar-tensor and $f(R)$ gravity. 
We have to  note that there is a difference in the exponent. 
In the case of  scalar-tensor gravity, the exponent is given by (\ref{grv9c}) but in case of 
 $f(R)$ gravity, by using (\ref{grv15}), it is 
\be
\label{grv9cF}
\hat h_{ij} \sim \left( \frac{t_E}{t_{E0}} \right)^{\tilde \eta\left( 1 - 3 \tilde \alpha \right)}
= \left( \frac{t}{t_0} \right)^{\tilde \eta\left( 1 - 3 \tilde \alpha \right)\left( - m + 2 \right)} 
\, ,
\ee
which is the relation of the solution in the Einstein and in the Jordan frame.

Here $\tilde\eta$, instead of (\ref{grv9b}), is given by
\be
\label{grv9bF}
\tilde\eta = {\tilde\eta}_\pm \equiv \frac{ 1 
\pm \sqrt{ 1 - \frac{16\tilde\alpha^2}{ \left( 1 - 3 \tilde\alpha \right)^2 }}}{2}\, .
\ee
We should  also note that $\tilde\alpha$ is defined according to (\ref{grv17}). 
For example, when $m\to -\infty$, we find $\alpha \to - m \to +\infty$ 
and $\tilde\alpha \to 3$
and therefore $\eta_\pm \to \frac{1}{2} \pm \frac{i}{6}$ and 
$\tilde \eta \to \frac{1}{2} \pm i \frac{\sqrt{5}}{4}$. 
Then the exponent is given by 
\be
\label{exp1}
\tilde \eta\left( 1 - 3 \tilde \alpha \right)\left( - m + 2 \right) 
\to \left( \frac{1}{2} \pm i \frac{\sqrt{5}}{4} \right) \left( - 8 \right) \left(-m \right) 
= - \left( 4 \pm 2i \sqrt{5} \right) \alpha \, .
\ee
The corresponding exponent in (\ref{grv9b}), for the scalar tensor gravity,  is
\be
\label{exp2}
\eta\left( 1 - 3 \alpha \right) \sim - \left( \frac{3}{2} \pm i\frac{\sqrt{7}}{2} \right) \alpha \, .
\ee
Then the  decreasing exponent  in the $f(R)$ gravity is much larger than the corresponding exponent 
in the scalar-tensor theory, the ratio is $8/3$. This means that if one is capable of fixing 
 the exponent by observations, we can distinguish the $f(R)$ gravity 
from the scalar-tensor theory. In this way, the evolution of the gravitons can be a tool to discriminate among competing cosmological models.

We stress again  that, although the phantom phase does not appear in the 
scalar-tensor theory, the phantom universe appears in  $f(R)$ gravity and in the cases $\frac{1}{2}<m<1$ and 
$\alpha<0$, $\hat h_{ij}$ is finite even near the Big Rip singularity. 
Therefore we can  easily distinguish the phantom universe from the quintessence  phase with 
$\alpha>1$ in the scalar-tensor theory and the $f(R)$ gravity because $\hat h_{ij}$ decreases in the 
quintessence phase. 

Some remarks are needed also regarding the Equation of State (EoS).
Because the FRW equations in the Einstein gravity coupled with perfect fluid are given by, 
\be
\label{JGRG11}
\rho_\mathrm{matter}=\frac{3}{\kappa^2}H^2 \, ,\quad 
p_\mathrm{matter}= - \frac{1}{\kappa^2}\left(3H^2 + 2\dot H\right)\, ,
\ee
it is convenient to define the effective EoS parameter in 
terms of the Hubble rate ${\displaystyle H\equiv \frac{\dot a}{a}}$ as follows, 
\be
\label{JGRG12}
w_\mathrm{eff}= - 1 - \frac{2\dot H}{3H^2} \, .
\ee
For the power law expanding universe in (\ref{grv5}), one finds ${\displaystyle H=\frac{\alpha}{t}}$ and 
therefore 
\be
\label{EoS1}
w_\mathrm{eff}= - 1 + \frac{2}{3\alpha} \, .
\ee
Then the limit of $\alpha\to + \infty$ in (\ref{exp1}) and (\ref{exp2}) corresponds to 
the limit of $w_\mathrm{eff} \to -1$, which is consistent with the $\Lambda$CDM model. 
We have also to note that there is no one-to-one correspondence between $\alpha$ 
and $m$ in the $f(R)$ gravity as in (\ref{grv11}). 
Besides,  the limits  $m\to -1$ and  $m\to 2 - 0$ give 
$\alpha \to + \infty$. 
In the limit of  $m\to 2 - 0$, we obtain the expression (\ref{exp2}) of the scalar-tensor 
theory even for the $f(R)$ gravity and therefore it becomes difficult to distinguish 
$f(R)$  from  scalar-tensor gravity. 

In the scalar-tensor theory, there is an enhancement by the factor $a(t)^2$ because the effective gravitational 
wave $h_{ij}$ is related to  $\hat h_{ij}$ by Eq.~(\ref{grv2}). 
Let us  now investigate what happens in $f(R)$ gravity. 
In the Einstein frame, we have a relation similar to Eq.~(\ref{grv2}), 
\be
\label{grv2FE}
h_{E\,ij} \left(\bm{x}, t_E \right) = \e^{i\bm{k}\cdot \bm{x}} a_E (t_E)^2 \hat h_{E\,ij}(t_E)\, . 
\ee
The scale factor $a_E$ in the Einstein frame is related with scale factor $a$ in the Jordan frame 
by $a_E^2 = f'(R) a^2$ as given by (\ref{grv12}). 
Furthermore Eq.~(\ref{grv12})  tells us that the gravitational wave $h_{ij}$ in the Jordan frame is related with 
the gravitational wave $h_{E\,ij}$ by $h_{ij} = f'(R)^{-1} h_{E,ij}$,
then we find 
\be
\label{grv2FE1}
h_{ij} \left(\bm{x}, t \right) = \e^{i\bm{k}\cdot \bm{x}} a (t)^2 
\hat h_{E\,ij}\left(t_E \left(t\right) \right)\, . 
\ee
Therefore there is an enhancement by a factor $a (t)^2$ but the situation is not changed with respect to that in the scalar tensor 
theory given in (\ref{grv12}).  As stated above, it is clear that the 
plane-wave amplitude evolution of the tensor graviton  strictly depends
on the background.

\section{Discusson and conclusions}
\label{cinque}

As we said, the only realistic approach to investigate  the  above interplay between the  cosmological background and  the graviton evolution is the study of the  stochastic background of gravitational waves.
 The   stochastic background coming from gravitons can be mainly related to
 two phenomena: 
 the (incoherent) superposition of gravitational waves
generated by  populations of astrophysical objects that cannot 
be   resolved \cite{Farmer:2003pa}, and the primordial 
background  of gravitational waves  coming from  early  epochs \cite{alebon, Bogdanos:2009tn}.
The stochastic background in 
described and characterized  by
a dimensionless spectrum 
\cite{Allen,Allen:1996gp,Maggiore:1999vm,Grishchuk:2000gh}, that is 

\begin{equation}
\Omega^A_\mathrm{GW}(\nu)=\frac{1}{\rho_{c}}\left(\frac{d\rho^A_\mathrm{GW}}{d\ln
\nu}\right)\, ,\qquad
\label{eq: spettro}
\mbox{with}\quad
\rho_{c}\equiv\frac{3H_{0}^{2}}{8\pi G} \, .
\end{equation}
Here $\Omega^A_\mathrm{GW}(\nu)$ is an dimensionless  density parameter,  $\rho_c$ is the today cosmic critical energy density, $H_0$ the today observed Hubble parameter, and
$d\rho_\mathrm{GW}$ is the energy density of  the
gravitational radiation  in the frequency range $\nu$ to
$\nu+d\nu$.
\begin{equation}
\rho_{\rm GW}=\int_0^{\infty}d\nu\,\tilde{\rho}_{\rm GW}(\nu)\,.
\end{equation}
  $\tilde{\rho}_{\rm GW}$
is the energy density of the gravitational waves  per unit frequency.
 The relation between $\Omega_{\rm GW}(\nu)$ and  the spectrum  $S_h(\nu)$ is \cite{Maggiore:1999vm,Allen:1987bk}
\begin{equation}
\Omega_{\rm GW}^A (\nu) = \left( \frac{4 \pi^2}{3H_0^2} \right) f^3 S_h^A (\nu)\, .
\label{eq22}
\end{equation}
The
energy density can be represented  as $h_{0}^2 \, \Omega_{\rm GW}(\nu)$ by
assuming  the dimensionless Hubble constant  $H_0=100\,h_{0}\,\rm{km\,
s^{-1}\,Mpc^{-1} }$. The  stochastic background energy density  is then
 written as
\be
\Omega_{\rm GW}^{A} \equiv  \Omega_{\rm GW}^{+} 
+ \Omega_{\rm GW}^{\times} + \Omega_{\rm GW}^S \, ,
\label{eq23}
\ee
with  $A=+,\times, S$.
The first two terms come directly from GR while the third term comes from alternative gravity and is due to the scalar modes. In the case of $f(R)$ gravity, a scalar field can be achieved from $f'(R)\rightarrow \Phi$  \cite{CapGW}.

The  stochastic background of gravitational waves is derived by the quantum fluctuations of 
zero-point energy. Such fluctuations are amplified in the early universe by the
large variations of gravity.  As discussed in 
Sec.~\ref{production},  this mechanism produces gravitational waves.  The mechanism 
is in agreement with  PLANCK data as well as with
almost exponential inflation and spectral index $\approx1$,
\cite{Ade:2015rim}.

It is worth noticing that  the inflationary scenario is related to
a natural mechanism giving rise to perturbations for
any field. Such a mechanism  
provides  a spectrum for relic scalar gravitons  which could constitute an important testbed for alternative  gravity
\cite{Allen,Grishchuk:2000gh,Vignolo,Corda1,Corda2}.  Introducing 
\begin{equation} 
 H_\mathrm{ds}^2=\frac{8\pi G\rho_\mathrm{ds}}{3c^{2}}\,,\qquad \rho_\mathrm{Planck}=\frac{c^{7}}{\hbar G^{2}}\, ,
\end{equation} 
that is,  the density at de Sitter (inflationary) epoch and the Planck density, respectively,  
the spectrum is given by
\begin{equation}
\Omega_\mathrm{GW}^{A}(\nu)=\frac{1}{\rho_{c}}\left(\frac{d\rho_\mathrm{GW}}{d\ln
\nu}\right)=\frac{\nu}{\rho_{c}}\left(\frac{d\rho_\mathrm{GW}}{d\nu}\right)=\frac{8}{9}\left(\frac{\rho_\mathrm{ds}}{\rho_\mathrm{Planck}}\right)\,.\label{eq: spettro gravitoni}
\end{equation} 
This
simple model  does not  consider the matter dominated era \cite{Cap2012}.
If we  include matter, we  have to take into
account  the equivalence epoch 
\cite{Allen:1987bk}
\begin{equation}
\Omega_\mathrm{GW}^{A}(f)=\frac{8}{9}\left(\frac{\rho_\mathrm{ds}}{\rho_\mathrm{Planck}}\right)(1+z_{eq})^{-1}\,,
\label{eq: spettro gravitoni redshiftato}
\end{equation} 
where $z_{eq}$ is the redshift at equivalence.
This means that  the range of frequencies of interest is 
$\nu>(1+z_{eq})^{1/2}H_{0}$. The   redshift $z_{eq}$ is a correction with respect to   the today 
Hubble parameter $H_{0}$. At lower frequencies, it is  \cite{Allen,Grishchuk:2000gh,Vignolo}
\begin{equation}
\Omega_\mathrm{GW}(\nu)\propto \nu^{-2}\, .
\label{eq: spettro basse frequenze}
\end{equation} 
Furthermore, it is worth noticing  
that Eqs. (\ref{eq: spettro gravitoni}) and 
(\ref{eq: spettro gravitoni redshiftato}) are  not correct at  any
frequency since they do not explicitly  depend on frequency. For gravitational waves with  frequencies less than $H_{0}$, the
energy density is  badly  defined because their wavelength is 
 larger than the Hubble size of the universe.  Similarly, at high frequencies, there is a maximal frequency above
which the spectrum rapidly drops to zero. We are assuming that the phase transition from
 inflation to radiation  is instantaneous. In realistic models, this phenomenon occurs over a given
 time scale related to a gravitational radiation maximal frequency.

However, $\Omega_\mathrm{GW}^{A}$ drops rapidly.  The 
cutoffs defined at low and high frequencies indicate
that the total energy density of the stochastic  gravitational waves has to be  finite.
These results have been quantitatively constrained taking into account  the
PLANCK experiment release. On the other hand, since the spectrum drops down as  
$\propto \nu^{-2}$ for  low frequencies, this means that,  at
LIGO-VIRGO  frequencies, one can get 
\begin{equation} 
\Omega_\mathrm{GW}^\mathrm{GR}(\nu)h_{100}^{2}<2 \times
10^{-6}\,,\quad 
\Omega_\mathrm{GW}^{S}(\nu)h_{100}^{2}<2.3\times
10^{-12}\, .
\label{eq: limite spettroGR}
\end{equation}

 The  corresponding strain   at
$\approx 100{\,\mathrm{Hz}}$, for VIRGO and LIGO at
maximum  sensitivity \cite{ligo,virgo} is an interesting quantity to be derived.  For this purpose,  one  can adopt  the  
characteristic amplitude equation for  a given component \cite{Maggiore:1999vm}. It is 
\begin{equation}
h_{A}(\nu)\simeq8.93\times10^{-19}\left(\frac{1\,\mathrm{Hz}}{\nu}\right)\sqrt{h_{100}^{2}\Omega_\mathrm{GW}(\nu)}\, .
\label{eq: legame ampiezza-spettro}
\end{equation} 
Finally,  we obtain 
\begin{equation}
h_{GR}(100\,\mathrm{Hz})<1.3\times 10^{-23}\,,\quad h_{s}(100\,\mathrm{Hz})<2\times 1.4 10^{-26}\,.
\label{eq: limite per lo strain}
\end{equation}
where GR and scalar modes are considered.

Because we have  sensitivities of the order  $10^{-22}$ for
the VIRGO and LIGO interferometers at $\approx100\,\mathrm{Hz}$, one needs to gain more or less
 three orders of magnitude. At smaller frequencies the
sensitivity of the VIRGO interferometer is of the order of
$10^{-21}$ at $\approx10\,\mathrm{Hz}$ and,  for the GR and scalar modes, it is 
\begin{equation}
h_{GR}(100\,\mathrm{Hz})<1.3 \times 10^{-22}\,,\quad h_{S}(100\,\mathrm{Hz})<1.4 \times 10^{-25}\,. 
\label{eq: limite per lo strain01}
\end{equation} 
These numerical results   point out  that tensor modes can be clearly distinguished by scalar modes. Furthermore, cosmological models can be characterized by the gravitational wave density parameter.

Other features can be put in evidence by the evolution of gravitons in a cosmological background. 
For example  a phantom universe cannot be realized in  scalar-tensor gravity  but can be 
realized in the $f(R)$ gravity. 
In general, it is difficult to distinguish the phantom universe with $w<-1$ from the quintessence universe 
$w>-1$,  if $w$ is almost $-1$. 

As discussed  after Eq.~(\ref{exp2}), in the phantom universe realized in the context of  $f(R)$ gravity, 
$\hat h_{ij}$ is finite but, in the quintessence phase, $\hat h_{ij}$ decreases. This fact, according to the analysis performed above, can have dramatic consequences  in the measurements of $h$ and of $\Omega_\mathrm{GW}$ and   therefore we could 
distinguish, in principle,  the phantom universe from the quintessence universe by the stochastic background of gravitational waves.

In the quintessence universe, realized  for $\alpha>1$, even if the expansion is identical for  the same $\alpha$, the exponents of $\hat h_{ij}$  are different for scalar-tensor  and  $f(R)$ gravity, as shown in (\ref{exp1}) and (\ref{exp2}). 
Then by observing the decreasing of  graviton in the long range, one can distinguish between  scalar tensor and $f(R)$ gravity.

The paradigm presented here can be adopted also for other models (e.g. Galileon, Gauss-Bonnet, etc.) as soon as one is capable of characterizing the source $\delta T_{\mu\nu}$ and a  graviton propagation equation  (\ref{eq:16}).  Clearly, such an equation depends on the conformal factor in (\ref{conformal}) that characterizes both the source and the model. 
From a genuine observational viewpoint, a main role is played by the upper limit (\ref{upper}) and the  amplitude (\ref{eq:27}). 
Both of them are related to  the Sachs-Wolfe effect and then to the capability of a fine determination of 
 cosmic microwave background temperature anisotropies. 
 
To conclude, we have shown that amplitudes of tensor gravitons and their evolution strictly depend on the
cosmological background. This background can be  tuned by a 
scalar field that specifically select  the theory of gravity.  Conversely, taking into account   that primordial vacuum fluctuations produce
stochastic gravitons, beside scalar perturbations, kinematical
distortions and so on, the initial amplitude of these gravitons  is a
function of the given theory of gravity.
Finally, the Sachs-Wolfe effect could
contribute to select a given    gravity theory  that  can be
consistently  matched with other observations. 

\section*{Acknowledgements}
This work is supported in part by the INFN sezione di Napoli, {\it iniziative specifiche} TEONGRAV and QGSKY (SC and MDL),
by MINECO (Spain)  project FIS2013-44881 (SDO) by the CSIC I-LINK1019 Project (SC, SN, and SDO). 
This work is  supported (in part) by 
MEXT KAKENHI Grant-in-Aid for Scientific Research on Innovative Areas ``Cosmic
Acceleration''  (No. 15H05890) (S.N.).
The  article is also based upon work from COST action CA15117 (CANTATA), 
supported by COST (European Cooperation in Science and Technology). M.D.L is supported by ERC Synergy
Grant "BlackHoleCam" Imaging the Event Horizon of Black Holes awarded by the ERC in 2013 (Grant No. 610058).


\end{document}